\documentclass[10pt,twocolumn,letterpaper,superscriptaddress]{revtex4}

\usepackage{multibbl}
\usepackage[pdftex]{graphicx}
\usepackage{amsmath,SIunits}
\usepackage[latin1]{inputenc}
\usepackage{latexsym,stmaryrd}
\usepackage[sort&compress]{natbib}
\usepackage[pdftex]{graphicx}
\usepackage{color}
\bibpunct{[}{]}{,}{n}{}{,}

\makeatletter
\newcommand*{\citenst}[2][]{%
  \begingroup
  \let\NAT@mbox=\mbox
  \let\@cite\NAT@citenum
  \let\NAT@space\NAT@spacechar
  \let\NAT@super@kern\relax
  \renewcommand\NAT@open{[}%
  \renewcommand\NAT@close{]}%
  \citep{#2}%
  \endgroup
}
\makeatother

\begin{document}

\title{Optomechanical coupling in the Anderson-localization regime}

\author{P. D. Garc\'{i}a}
\email{david.garcia@icn2.cat}
\affiliation{Catalan Institute of Nanoscience and Nanotechnology (ICN2), CSIC and The Barcelona Institute of Science and Technology, Campus UAB, Bellaterra, 08193 Barcelona, Spain}
\author{R. Bericat-Vadell}
\affiliation{Catalan Institute of Nanoscience and Nanotechnology (ICN2), CSIC and The Barcelona Institute of Science and Technology, Campus UAB, Bellaterra, 08193 Barcelona, Spain}
\affiliation{Dept. de F\'{i}sica, Universitat Autonoma de Barcelona, 08193 Bellaterra, Spain}
\author{G. Arregui}
\affiliation{Catalan Institute of Nanoscience and Nanotechnology (ICN2), CSIC and The Barcelona Institute of Science and Technology, Campus UAB, Bellaterra, 08193 Barcelona, Spain}
\affiliation{Dept. de F\'{i}sica, Universitat Autonoma de Barcelona, 08193 Bellaterra, Spain}
\author{D. Navarro-Urrios}
\affiliation{Catalan Institute of Nanoscience and Nanotechnology (ICN2), CSIC and The Barcelona Institute of Science and Technology, Campus UAB, Bellaterra, 08193 Barcelona, Spain}
\author{M. Colombano}
\affiliation{Catalan Institute of Nanoscience and Nanotechnology (ICN2), CSIC and The Barcelona Institute of Science and Technology, Campus UAB, Bellaterra, 08193 Barcelona, Spain}
\affiliation{Dept. de F\'{i}sica, Universitat Autonoma de Barcelona, 08193 Bellaterra, Spain}
\author{F. Alzina}
\affiliation{Catalan Institute of Nanoscience and Nanotechnology (ICN2), CSIC and The Barcelona Institute of Science and Technology, Campus UAB, Bellaterra, 08193 Barcelona, Spain}
\author{C. M. Sotomayor-Torres}
\homepage{http://www.icn.cat/~p2n/}
\affiliation{Catalan Institute of Nanoscience and Nanotechnology (ICN2), CSIC and The Barcelona Institute of Science and Technology, Campus UAB, Bellaterra, 08193 Barcelona, Spain}
\affiliation{ICREA - Instituci\'o Catalana de Recerca i Estudis Avan\c{c}ats, 08010 Barcelona, Spain}

\small

\begin{abstract}
Optomechanical crystals, purposely designed and fabricated semiconductor nanostructures, are used to enhance the coupling between the electromagnetic field and the mechanical vibrations of matter at the nanoscale.\ However, in real optomechanical crystals, imperfections open extra channels where the transfer of energy is lost, reducing the optomechanical coupling efficiency.\ Here, we quantify the role of disorder in a paradigmatic one-dimensional optomechanical crystal with full phononic and photonic bandgaps.\ We show how disorder can be exploited as a resource to enhance the optomechanical coupling beyond engineered structures, thus providing a new toolset for optomechanics.
\end{abstract}


\maketitle

The coupling of electromagnetic radiation to mechanical vibrations is at the heart of solid-state quantum photonics~\cite{QP} while phonon transport at different frequencies governs crucial physical phenomena ranging from thermal conductivity~\cite{Ibach} to the sensitivity of nano-electromechanical resonators~\cite{MEMS}.\ This coupling can be controlled very efficiently by optomechanical crystals~\cite{Eichenfield}, like nanobeams~\cite{Painter,Lonkar,Jordi}, where the electromagnetic field and the mechanical displacement can be colocalized spatially within defect modes engineered in the structure.\ In the ideal case, i.e., in absence of imperfections, the coupling efficiency between phonons and photons is just limited by intrinsic losses such as thermoelastic damping~\cite{Zener}.\ However, further energy dissipation is imposed by unavoidable fabrication imperfections which open extra leaky channels dramatically reducing the optomechanical coupling efficiency.\ Phonons are particularly sensitive to fabrication imperfections which can break the symmetry allowing the coupling among phononic Bloch modes with different symmetry~\cite{Chan1} and reducing the optomechanical coupling efficiency.\ Since disorder is considered detrimental, efforts are usually oriented on minimizing it~\cite{Chan2,topology}.\ Here, we propose a different strategy focused on exploiting disorder as a resource.

\begin{figure}[t!]
  \includegraphics[width=\columnwidth]{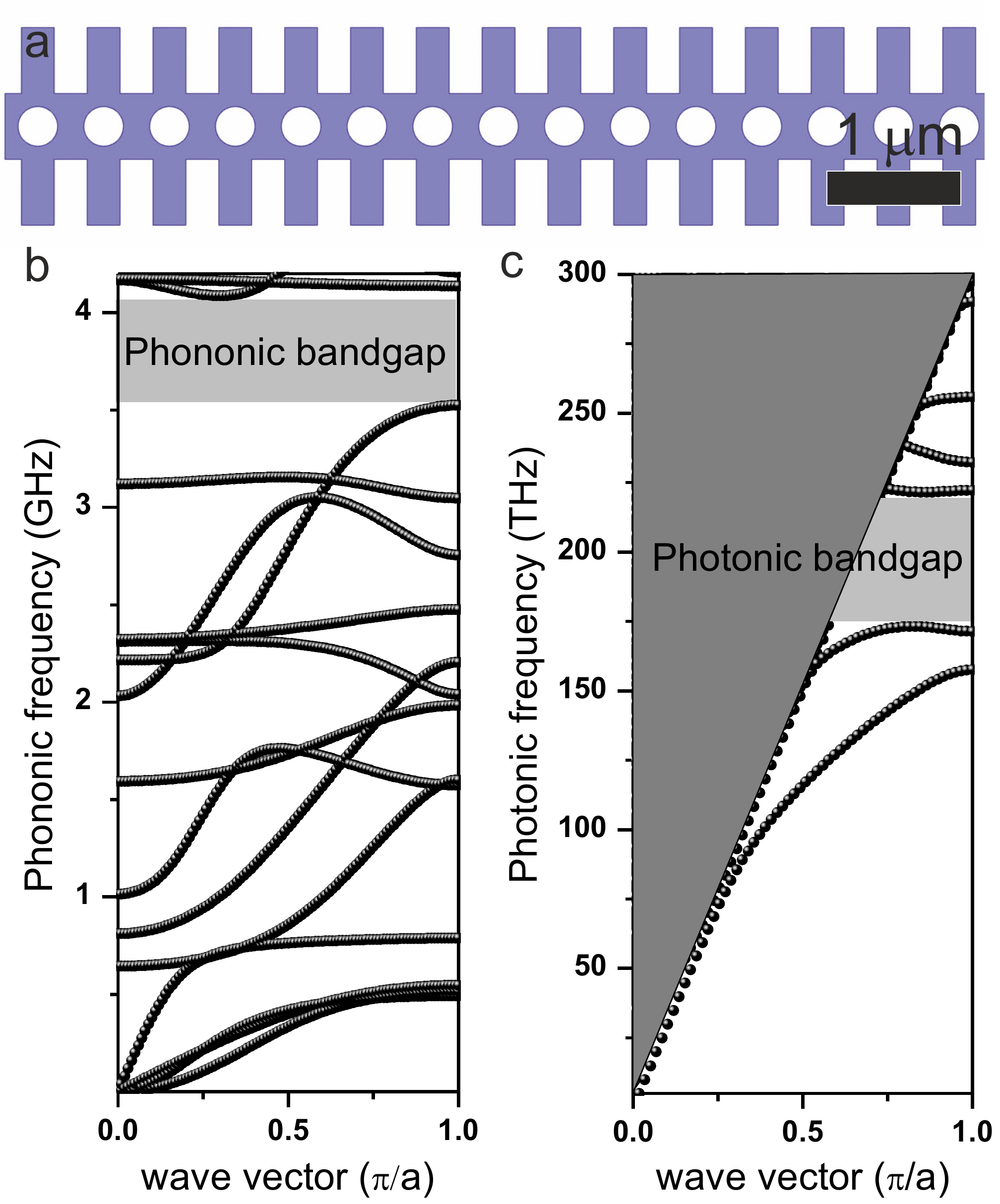}
      \caption{ \label{1} \textbf{A one-dimensional optomechanical crystal: the nanobeam.}
      The nanobeam used in our calculations is shown in (a).\ Holes are created with a period of $a=500\,\text{nm}$ and a radius $\text{r}=0.3a$ in a beam of width $\text{w}=a$ and thickness $\text{t}=0.44a$.\ Wings are incorporated with length $\text{l}=3a$ and a width $\text{d}=0.5a$ which opens a full phononic (b) and photonic bandgap (c) in the GHz and hundreds of THz frequency range of the spectrum, respectively.}
    \end{figure}

\begin{figure}
  \includegraphics[width=\columnwidth]{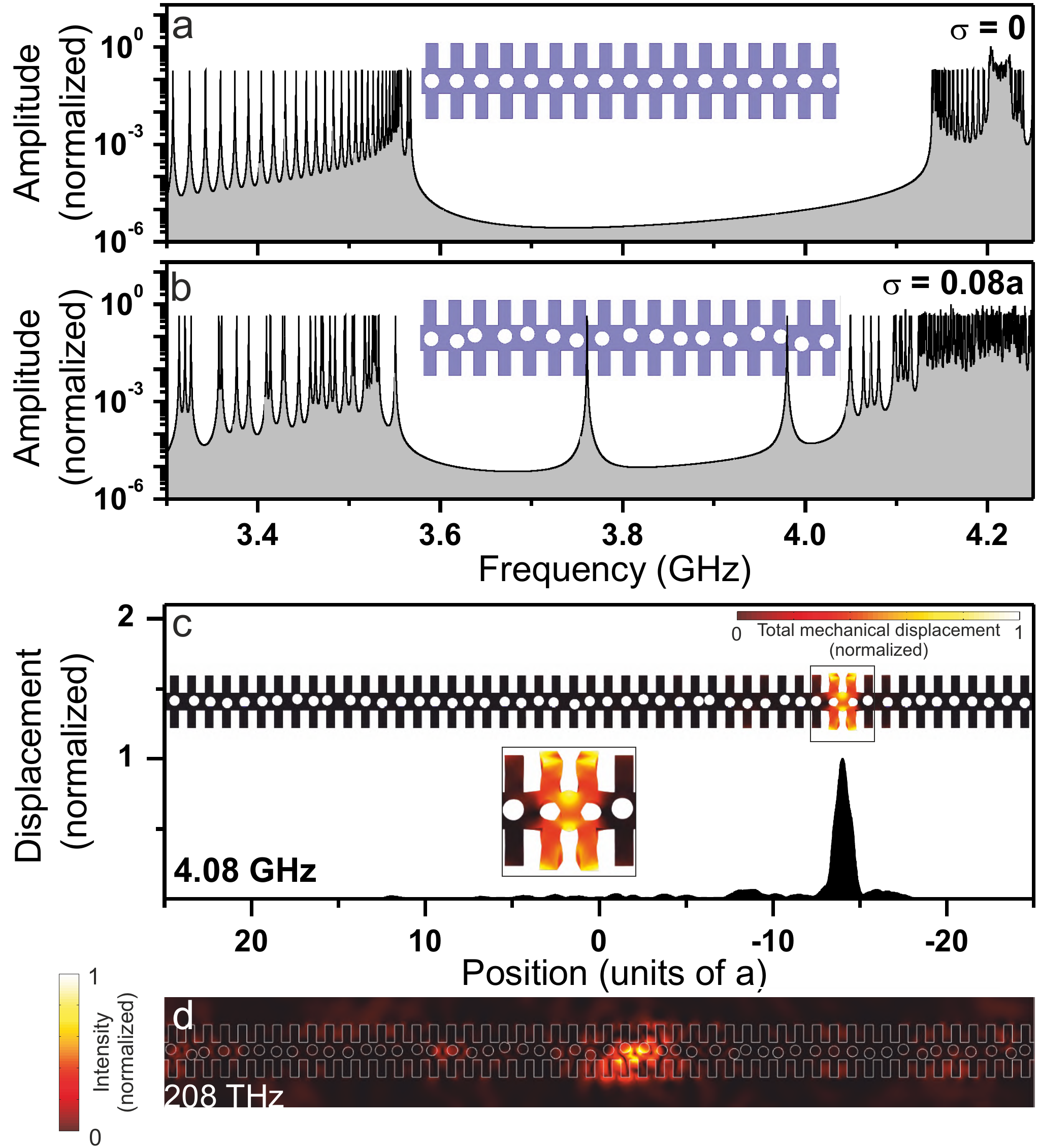}
    \caption{ \label{2} \textbf{A mechanical Lifshitz tail.} (a) Calculated mechanical eigenmodes of a nanobeam with $\sigma = 0$ (ideal structure) and a total length of $\text{L}=100a$ showing a clear bandgap in the GHz frequency range.\ (b) Mechanical eigenmodes calculated for a disordered nanobeam where the positions of the holes are randomized normally with a standard deviation $\sigma = 0.08a$.\ Examples of the total mechanical displacement corresponding to a disorder-induced mechanical mode (c) and a disorder-induced photonic mode (d) localized within the same perturbed structure.}
\end{figure}

When a crystal is structurally perturbed such as a crystalline solid, a photonic crystal or an optomechanical crystal, the ideally propagating Bloch modes undergo random multiple scattering and become sensitive to imperfections specially at the edges of the band gap~\cite{John}.\ In the fully three-dimensional case~\cite{Sheng}, a phase transition occurs when disorder is sufficiently high and the modes become localized states, decaying exponentially when ensemble-averaged with a characteristic length scale called the localization length, $\xi$.\ In the one-dimensional structure analyzed here, the crossover from extended to localized modes occurs at $\xi \leq \text{L}$, where L is the total length of the structure~\cite{Sheng}.\ These disorder-induced narrow resonances populate the spectrum around the band edges forming a band of localized modes known as Lifshitz tail~\cite{Lifshitz}, which broadens with increasing disorder.\ This type of confinement, also known as Anderson localization, was proposed in the context of electronic transport to explain the metal-insulator phase transition induced by structural imperfections in a semiconductor~\cite{Anderson}.

In this Letter, we explore the role of disorder on phonon propagation and on the optomechanical coupling.\ We use a commercially available finite-element method solver to explore phonon Anderson localization in the nanobeam structure plotted in Fig.~\ref{1}a.\ We calculate the total deformation vector of the structure, \textbf{u}, considering continuum linear mechanics theory~\cite{Continuus}.\ In detail, we perform an eigenfrequency analysis to obtain the mechanical eigenmodes of the system and a frequency domain analysis to obtain the steady-state deformation of the structure after a force applied at a frequency $\omega_{F}$ typically at the center of the structure.\ The eigenmodes of this particular design are distributed in bands with a full gap opened for GHz phonons and hundreds of THz photons~\cite{Oudich,Jordi}, as shown in Fig.~\ref{1}b and \ref{1}c.\ The structure is obtained by creating holes with a period of $a=500\,\text{nm}$ and radius $\text{r}=0.3a$ on a silicon membrane beam with thickness $\text{t}=0.44a$ and width $\text{w}=a$.\ The total length of the simulated structure is $\text{L}=100a$.\ Fig.~\ref{2}a plots the mechanical spectrum calculated for the (finite) perfect structure, revealing a clear phononic gap which expands from about 3.6 GHz to 4.1 GHz.\ The quality factors of the mechanical modes are obtained by including thermoelastic damping in our full three-dimensional simulations~\cite{Thermoelastic}.\ Disorder is introduced by randomizing the positions of the holes by $\Delta \text{\textbf{r}}$, normally distributed around their ideal position with a standard deviation $\sigma = \sqrt{\langle \Delta \text{\textbf{r}}^2\rangle}$, where $\langle \Delta \text{\textbf{r}} \rangle = 0$.\ The brackets here indicate the expectation value.\ Fig.~\ref{2}b shows how the gap is populated with disorder-induced resonances which are strongly localized along the crystal perturbed by $\sigma=0.08a$.\ Fig.~\ref{2}c plots an example of an Anderson-localized mechanical mode with an effective volume of $\text{V}_{\text{eff}}=0.1\,\micro\text{m}^3$, which represents half of the effective volume of a \textit{perfect} engineered mechanical cavity~\cite{Oudich}.\ Although we focus here on phonon localization, the photonic counterpart also becomes localized as shown in Fig.~\ref{2}d for $\sigma=0.08a$.

\begin{figure}
  \includegraphics[width=\columnwidth]{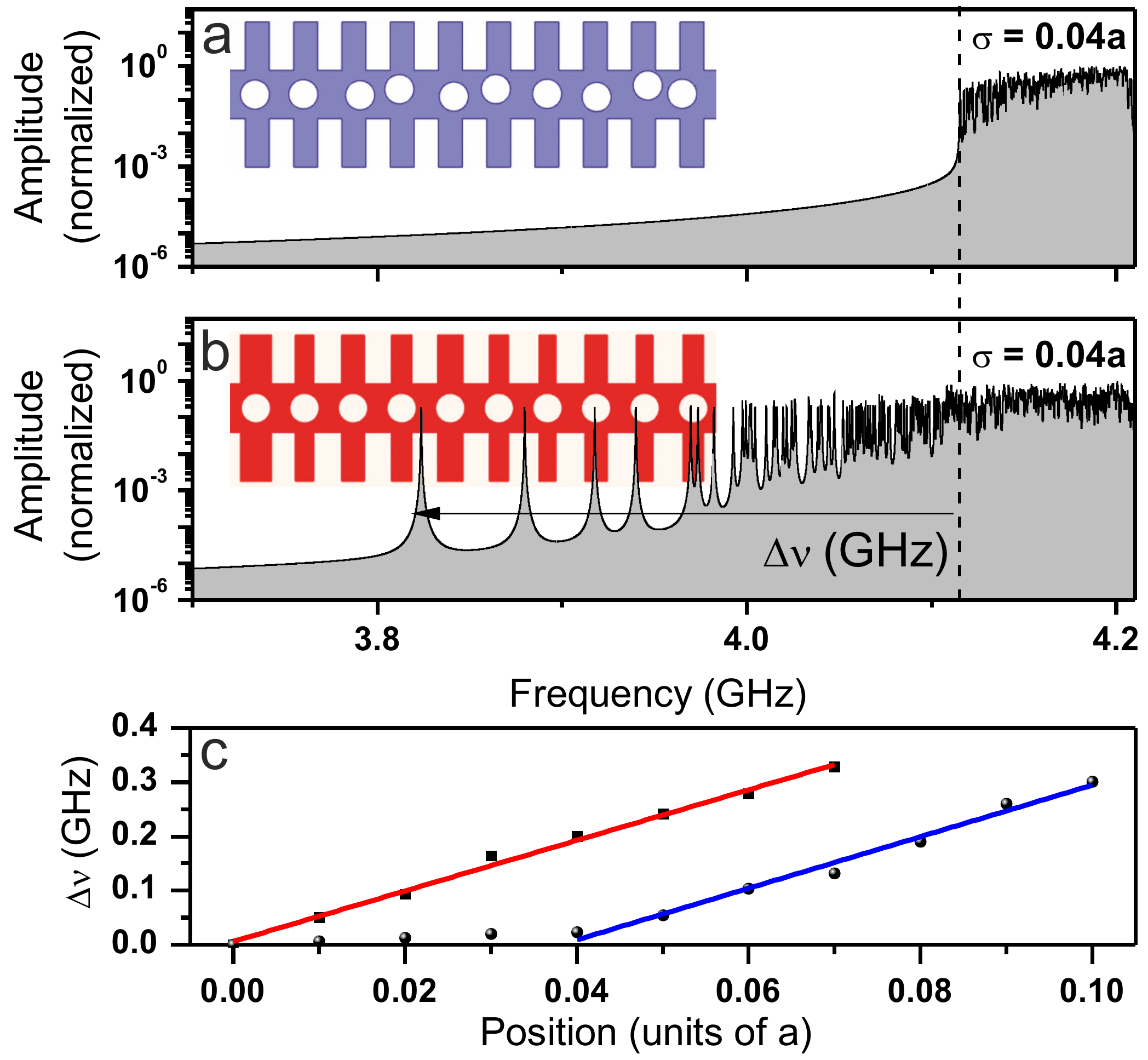}
    \caption{ \label{3} \textbf{Perturbation of holes and wings} (a) Mechanical eigenmodes calculated for a disordered nanobeam where the positions of the holes are randomized normally with a standard deviation $\sigma = 0.04a$.\ (b) Mechanical eigenmodes calculated for a disordered nanobeam where the width of the wings are randomized normally with a standard deviation $\sigma = 0.04a$.\ The Lifshitz tail width corresponding to perturbation of the wings width (squares) and the hole position (circles).}
\end{figure}

The  mechanical properties of this structure are even more sensitive to the perturbation of the wings of the unit lattice.\ Fig.~\ref{3} shows how the gap is populated with disorder-induced resonances when perturbing either the position of the holes (a) or the width of the wings (b) the same amount, $\sigma = 0.04a$.\ We remark that the optical and the mechanical properties are affected differently by the structural parameters of the unit cell, as explained in Ref~\citenst{Jordi} and confirmed here by our calculations.\ For example, the electromagnetic field is much more sensitive to the positions of the holes while the mechanical vibrations are more affected by the wings and localized when perturbing their position just by $\sigma > 0.01a$, as plotted in Fig.~\ref{3}c.\ We attribute this to the fact that the wings distribute the mass anisotropically within the unit cell thus providing the necessary interference to open a full mechanical gap~\cite{Oudich}.\ This different response opens the exciting possibility to engineer their colocalization independently to enhance the optomechanical coupling rate even further.\

\begin{figure}
  \includegraphics[width=\columnwidth]{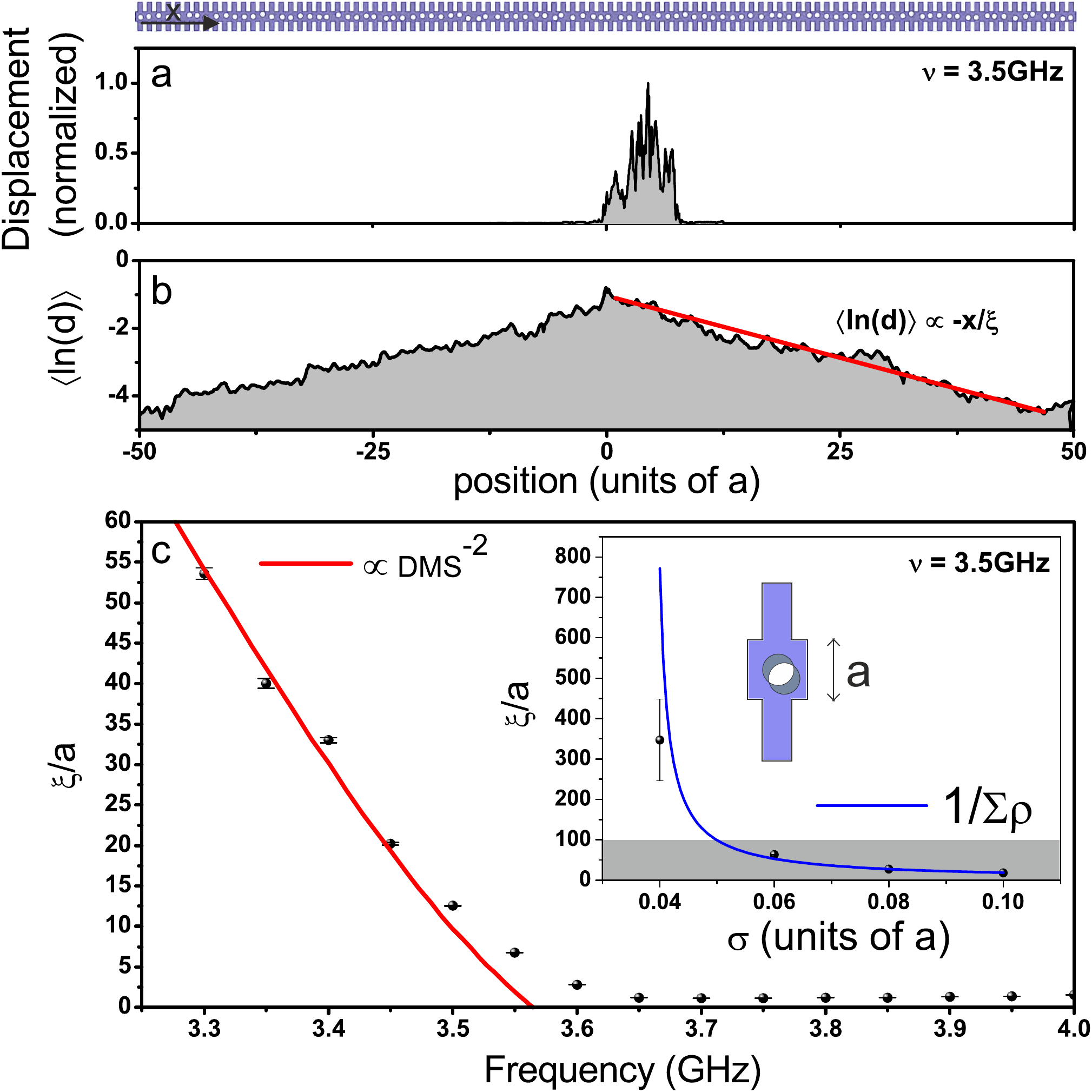}
    \caption{ \label{4} \textbf{Phononic localization length.} (a) Total mechanical displacement resulting from a force applied at the center of the nanobeam along the x direction with a frequency $\omega_{F}=3.5~\text{GHz}$.\ (b) Total mechanical displacement after ensemble-averaging over 10 realizations of disorder at the same frequency.\ The slope yields the inverse of the localization length, $\xi$.\ (c) Dispersion of the $\xi$ for $\sigma = 0.08a$ where the red line plots the density of mechanical states (DMS).\ The inset to the figure shows the dependence of $\xi$ vs. disorder for a fixed frequency of $\omega_{F}=3.5~\text{GHz}$.}
\end{figure}

The figure of merit of an Anderson-localized cavity is the localization length, $\xi$, its ensemble-averaged exponential spatial decay.\ $\xi$ is a key parameter in the Anderson localization regime, i.e., in photonics it determines the degree of confinement and the coupling strength between a quantum light emitter to a localized mode~\cite{Smolka2011,Tyrrestrup} as well as the efficiency of a random laser working in the Anderson-localization regime~\cite{Jin}.\ The localization length sets the crossover for strong disorder in low dimensional systems as $\xi \leq \text{L}$.\ To obtain $\xi$ for phonons, we calculate the steady-state total mechanical displacement resulting from a force, $\vec{F} = F\vec{i}$, applied at the center of the structure at a frequency $\omega_{F}$ with an arbitrary amplitude.\ Three lattice units of perfect reflectionless absorbers are added at both terminations to mimic an open system and to avoid unwanted reflections.\ As plotted in Fig.~\ref{4}a for $\omega_{F}=3.5\,\text{GHz}$, this mechanical displacement is strongly fluctuating along the structure but decays exponentially after ensemble-averaging over ten configurations as shown in Fig.~\ref{4}b.\ This exponential decay is a fingerprint of Anderson localization~\cite{Anderson_universality} which is non trivial to observe in a periodic nanostructure~\cite{Lalanne_2}.\ Fig.~\ref{4}c plots the strongly dispersive $\xi$ for $\sigma = 0.08a$, which decreases well below the sample length, $\text{L} = 100a$, down to a lattice unit within the phononic gap.\ This strong dispersion is due to a delicate interplay between order and disorder which has already been explored in other perturbed periodic nanostructures where only the photonic~\cite{Lalanne_2,Garcia2010} or the phononic~\cite{Modinos} localization were independently analyzed.\ A fully statistical analysis of the phase space of this particular physical system, which is certainly rich, e.g., frequency, amount of disorder and type of disorder, is necessary to maximize the optomechanical coupling induced by disorder and it is out of the scope of the present manuscript.

In the low-frequency range in Fig.~\ref{4}c, $\xi$ scales as $\text{DMS}^{-2}$, where $\text{DMS}$ is the density of mechanical states calculated as $[\partial \omega(k)/\partial k]^{-1}$ and $\omega(k)$ the phononic dispersion relation of the ideal structure plotted in Fig.~\ref{1}b.\ This scaling has been observed in photonic-crystal waveguides~\cite{Garcia2010} and in three-dimensional photonic crystals~\cite{PRBscattering} and it is attributed to a modified scattering cross section in a perturbed periodic structure.\ To explain this scaling, there are basically three approximations to take into account.\ First, in a one-dimensional single-mode structure, the localization length equals the scattering mean free path~\cite{Sheng,Beenakker}, i.e., $\xi \approx \ell_\text{s}$.\ In addition, as our system and our calculations are fully three dimensional, the scattering mean free path can be expressed $\ell_\text{s} = 1/\rho_\text{s} \Sigma$, where $\rho_\text{s}$ is the number density of scatterers and $\Sigma$ is the scattering cross section~\cite{Sheng}.\ Finally, two separate mechanisms determine $\Sigma$ in a periodic structure: how the Bloch mode couples to the scatterer and how the scatterer radiates the scattered wave.\ While the former is described by the density of states along the incident wavevector $k$ \cite{Mcphedran}, the latter reduces also to the density of states when considering only in-plane scattering~\cite{Hughes_purcell} and gives rise to the scaling of $\xi$ with $\text{DMS}^{-2}$ shown here.\ In the inset to Fig.~\ref{4}c, we plot $\xi$ vs. amount of disorder at a frequency $\omega_{F}=3.5\,\text{GHz}$.\ Here, we assume the difference between the perturbed and the ideal hole~\cite{area} as the scattering source yielding a scattering cross section $\Sigma = (0.60 \pm 0.03)a^2 = (6 \pm 0.1)r^2$, which is comparable to the area of a \textit{full hole} and shows a dramatically enhanced scattering response at this particular frequency.

\begin{figure}
  \includegraphics[width=\columnwidth]{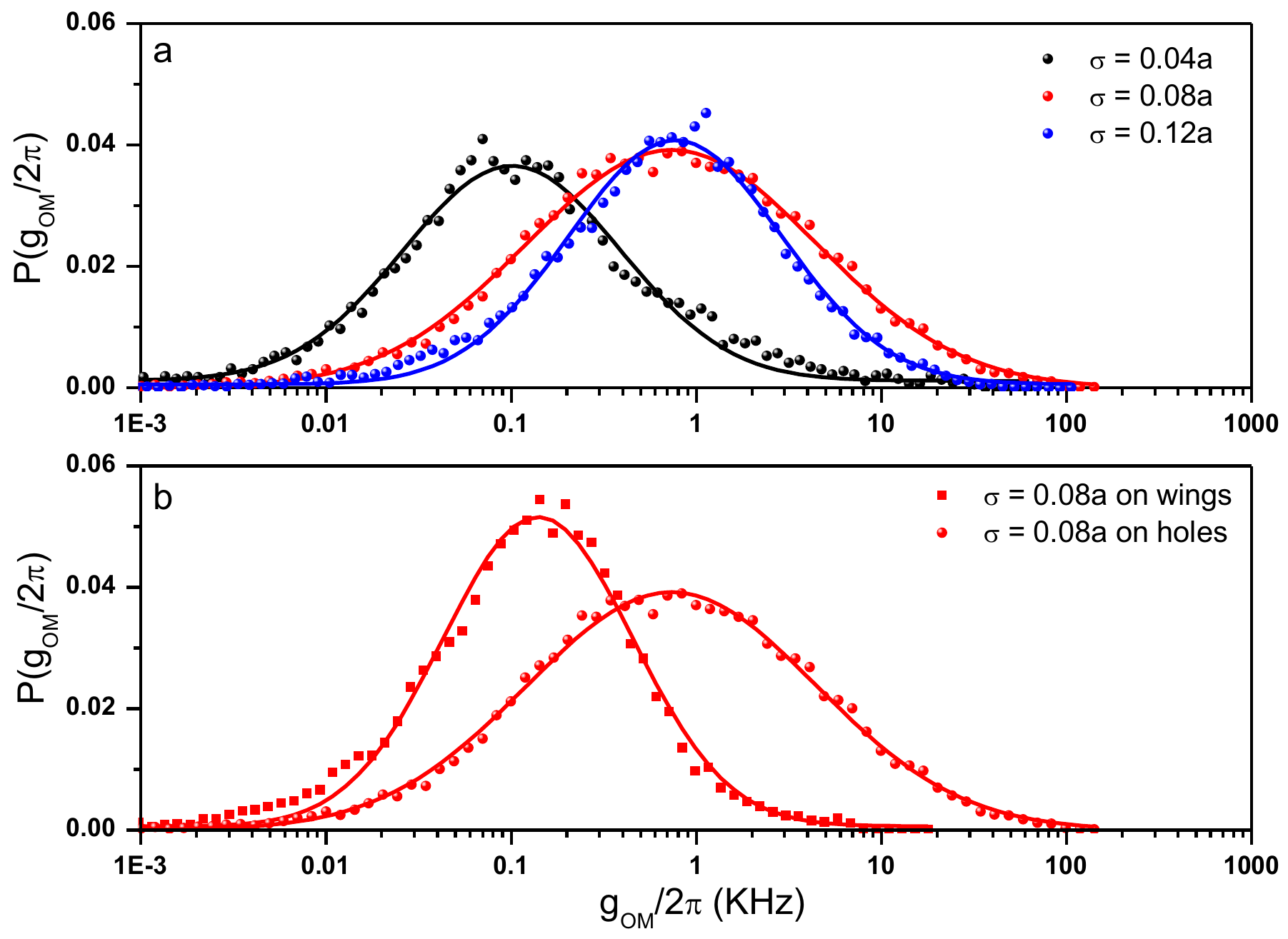}
    \caption{ \label{5} \textbf{Optomechanical coupling in the Anderson localization regime.}
    (a) Probability distribution of the vacuum optomechanical coupling rate, $\text{g}_\text{OM}/2\pi$, calculated between all the Anderson-localized mechanical and photonic modes found in ten nanobeams randomized with a standard deviation $\sigma=0.04a$ (black dots),  $\sigma=0.08a$ (red dots) and $\sigma=0.12a$ (blue dots).\ We fit them with a log-normal distribution.\ (b) $\text{g}_\text{OM}/2\pi$ calculated when perturbing the width of the wings by $\sigma=0.08a$ (red squares), compared to the coupling calculated when perturbing the holes by the same amount (red dots).}
\end{figure}

Finally, we calculate the vacuum optomechanical coupling rate $\text{g}_\text{OM}/2\pi$ between all the photonic and phononic Anderson-localized modes found in a set of structures perturbed by different amounts of disorder.\ Details of the calculation of $\text{g}_\text{OM}/2\pi$ can be found elsewhere~\cite{Chan2}.\ Fig.~\ref{5}a plots the probability distribution of the calculated $\text{g}_\text{OM}/2\pi$ for ten structures perturbed by $\sigma=0.04a$ (black dots), $\sigma=0.08a$ (red dots) and $\sigma=0.12a$ (blue dots).\ Deep in the localization regime, many variables deviate from a normal distribution showing heavy-tailed distributions such as, e.g., the quality factor and the volume of localized modes \cite{Tyrrestrup}, the transmission intensity \cite{Rossum} or the conductance fluctuations \cite{Beenakker}.\ In this case, we also observe a log-normal distribution of $\text{g}_\text{OM}/2\pi$ in the localization regime.\ The mean value of the log-normal distribution increases with disorder, while the variance has a maximum value for $\sigma=0.08a$ and then decreases.\ We attribute this dependence on disorder to a non-trivial interplay between the localization length and the photonic leakage out-of-the structure.\ Up to $\sigma=0.08a$, the localization length is comparable to the total length of the structure (see inset to Fig.~\ref{4}c), thus giving rise to extended leaky modes with poor coupling rates - in the range of 0.1 kHz.\ With increasing disorder, the localization length decreases giving rise to strongly confined modes within the structure with larger coupling rates.\ For larger perturbation, however, also the leakage of photonic modes increases which reducing both the photonic confinement and the optomechanical coupling.\ The maximum rates are calculated in the 100s of kHz range and correspond to strongly overlapping photonic and phononic localized modes, which are at experimental reach as we have explored only ten structures.\ We compare these values with the coupling rates calculated for an engineered tapered cavity in the same structure~\cite{Oudich}, which offers a maximum coupling rate in the 10s of kHz~\cite{Oudich,Jordi}.\ Although it is possible to improve significantly this value by band engineering, the coupling between \textit{perfect} bare cavity modes is significantly lower than the maximum values shown here.\ For completeness , Fig.~\ref{5}b plots the probability distribution of the coupling rate calculated when perturbing the width of the wings and the position of the holes by $\sigma=0.08a$.

In conclusion, we present a numerical analysis of Anderson localization in optomechanical crystals with particular attention to phonon localization which opens the possibility to study the role of polarization in Anderson localization~\cite{Skipetrov}.\ Our calculations demonstrate an alternative route to explore optomechanical coupling at the nanoscale well beyond the state-of-the art where imperfections play a central role.\ Here, we calculate the vacuum optomechanical coupling rate between disorder-induced modes that overcome the coupling rate of engineered cavity modes.\ Controlling Anderson localization in these nanostructures can also bring innovative solutions to open issues in a broad range of scientific disciplines, e.g., slowing down the dephasing time scale of spin qubits~\cite{Spintronics} or even for thermal insulation~\cite{Therm} at very low temperatures (mK).

\textbf{Acknowledgements}

This work was supported by the Spanish Severo Ochoa Excellence program and the MINECO project PHENTOM (FIS2015-70862-P).\ P.~D. Garc\'{i}a and D. Navarro-Urrios gratefully acknowledge the support of a Marie Sk\l{}odowska-Curie Individual fellowship (GAT-701590-1), a Marie  and a Ramon y Cajal fellowship (RYC-2014-15392), respectively.


\end{document}